\documentstyle[aps,prl,multicol,epsfig]{revtex}
\newcommand{\bq}{\begin{equation}}
\newcommand{\eq}{\end{equation}}
\newcommand{\bqa}{\begin{eqnarray}}
\newcommand{\eqa}{\end{eqnarray}}
\newcommand{\nn}{\nonumber \\}

\begin{document}
\draft
\title{Origin of the Quasiparticle Peaks of Spectral Functions in High $T_c$ Cuprates}

\author{Sung-Sik Lee and Sung-Ho Suck Salk$^a$}
\address{Department of Physics, Pohang University of Science and Technology,\\
Pohang, Kyoungbuk, Korea 790-784\\
$^a$ Korea Institute of Advanced Studies, Seoul 130-012, Korea\\}
\date{\today}

\maketitle

\begin{abstract}
Based on the SU(2) slave-boson approach to the t-J Hamiltonian, we examine the cause of the sharp peaks('quasiparticle' peaks) in the observed spectral functions in high $T_c$ cuprates. 
The computed results reveal that the spectral weight of the sharp peaks increases with hole doping rate in agreement with observation. 
It is shown that the observed sharp peaks are attributed to the enhancement of spinon pairing(spin singlet pair formation) by the presence of holon pair bosons in the superconducting state.
\end{abstract}
\begin{multicols}{2}

\newpage
Recently we reported a study of phase diagram\cite{SALK} involving holon pair condensation for high $T_c$ cuprates based on an improved approach of the SU(2) slave-boson theory\cite{WEN} over a previous study\cite{GIMM} of the U(1) slave-boson theory that we recently made.
In this approach, both the spinon and holon degrees of freedom are introduced into the Heisenberg exchange term in the t-J Hamiltonian, by considering the possibility of on-site charge fluctuations which arise as a result of site to site electron(and thus holon) hopping for the quantum systems of hole doped high $T_c$ cuprates.
Unlike the SU(2) theory, the phase fluctuation effects of order parameters are not taken into account in the U(1) mean field approach. 
Thus it is of great interest to study how the phase fluctuations affect the observed spectral functions by applying the SU(2) theory with the above considerations.
Currently there exists a lack of understanding the microscopic cause of the sharp peaks('quasiparticle' peaks) in the ARPES (angle resolved photoemission spectroscopy)\cite{SHEN}\cite{NORMAN}. 
In the present study, using the improved approach of the SU(2) slave-boson theory\cite{SALK}\cite{WEN} we evaluate one particle spectral functions for the normal and superconducting states and focus on the cause of the sharp quasiparticle peaks which appear in the superconducting state.
In addition we examine the role of phase fluctuations of the spinon pairing order parameters on the spectral functions based on the SU(2) theory.

In the slave-boson representation\cite{KOTLIAR}\cite{WEN}, the electron annihilation operator of spin $\sigma$, $c_{\sigma}$ can be written as a composite of spinon and holon operators. 
That is, $c_{\sigma}  =  b^\dagger f_{\sigma}$ in the U(1) representation and 
$c_{\alpha}  = \frac{1}{\sqrt{2}} h^\dagger \psi_{\alpha}$ in the SU(2) theory with $\alpha=1,2$,
where $f_{\sigma}$($b$) is the spinon(holon) annihilation operator in the U(1) theory,
and $\psi_1=\left( \begin{array}{c} f_1 \\ f_2^\dagger \end{array} \right)$ $\left(\psi_2 = \left( \begin{array}{c} f_2 \\ -f_1^\dagger \end{array} \right) \right)$ and $h = \left( \begin{array}{c} b_{1} \\ b_{2} \end{array} \right)$ are the doublets of spinon and holon annihilation operators respectively in the SU(2) theory.

Introducing Hubbard Stratonovich transformations for direct, exchange and pairing channels and a subsequent saddle point approximation, the t-J Hamiltonian is decomposed into the spinon sector, $H^f$ and the holon sector, $H^b$\cite{SALK},
\bqa
\lefteqn{H^f  =  -\frac{J(1-\delta)^2}{2} \sum_{<i,j>} \Bigl[ \Delta_{ij}^{f*} (f_{1j}f_{2i}-f_{2j}f_{1i}) + c.c. \Bigr] }\nn
&& - \frac{J(1-\delta)^2}{4} \sum_{<i,j>} \Bigl[ \chi_{ij} (f_{\sigma i}^{\dagger}f_{\sigma j}) + c.c. \Bigr], \nn
\lefteqn{H^b  =  -\frac{t}{2} \sum_{<i,j>} \Bigl[ \chi_{ij}(b_{1i}^{\dagger}b_{1j} - b_{2j}^{\dagger}b_{2i}) -\Delta^f_{ij} (b_{1j}^{\dagger}b_{2i} + b_{1i}^{\dagger}b_{2j})\Bigr] - c.c. }\nn
&& - \sum_{<i,j>,\alpha,\beta} \frac{J}{2}|\Delta^f_{ij}|^2 \Bigl[ \Delta_{ij;\alpha \beta }^{b*} (b_{\alpha i}b_{\beta j}) + c.c. \Bigr] - \sum_{i,\alpha} \mu_i b_{\alpha i}^\dagger b_{\beta j},
\eqa
where $\chi_{ij}= < f_{\sigma j}^{\dagger}f_{\sigma i} + \frac{2t}{J(1-\delta)^2} (b_{1j}^{\dagger}b_{1i} - b_{2i}^\dagger b_{2j} )>$ is hopping order parameter, $\Delta_{ij}^{f}=< f_{1j}f_{2i}-f_{2j}f_{1i} >$, spinon pairing order parameter, $\Delta_{ij;\alpha\beta}^{b} = <b_{i\alpha}b_{\beta j}>$, holon pairing order parameter, and $\mu_i$, the effective chemical potential.
With the uniform hopping order parameter, $\chi_{ij}=\chi$, the d-wave spinon pairing order parameter, $ \Delta_{ij}^{f}=\pm \Delta_f$ with the sign $+(-)$ for the nearest neighbor link parallel to $\hat x$ ($\hat y$) and the s-wave holon pairing order parameter, $\Delta_{ij;\alpha \beta}^{b}=\Delta_b(\delta_{\alpha,1}\delta_{\beta,1} - \delta_{\alpha,2}\delta_{\beta,2} )$, the quasiparticle energy for spinon is given by\cite{SALK}
\bqa
E_{k}^{f}  & = &  \sqrt{(\epsilon_{k}^{f})^{2} + ( \Delta_f^{'})^2},
\label{eq:spinon_energy}
\eqa
where the spinon single particle energy is given by,
\bqa
\epsilon_{k}^{f} & = & -\frac{J(1-\delta)^2}{2} \chi (\cos k_x + \cos k_y),
\label{eq:spinon_energy_single}
\eqa
and the spinon pairing gap,
\bqa
\Delta_f^{'} & = & J (1-\delta)^2 \Delta_f (\cos k_x - \cos k_y).
\label{eq:spinon_energy_gap}
\eqa
The single particle(electron) propagator of interest is given by a convolution integral of spinon and holon propagators in the momentum space\cite{WEN},
\bqa
G_{\alpha\beta}({\bf k},\omega) & = & i\int \frac{d{\bf k}^{'} d\omega^{'}}{(2\pi)^3} G^f_{\alpha\beta}({\bf k}+{\bf k}^{'},\omega+\omega^{'}) G^b({\bf k}^{'},\omega^{'}) \\
&& \hspace{1cm}\mbox{in the U(1) theory, and} \nn
G_{ \alpha \beta}({\bf k},\omega) & = & \frac{i}{2} \int \frac{d{\bf k}^{'} d\omega^{'}}{(2\pi)^3} \Bigl[ \sum_{l,m}G^f_{ \alpha \beta l m }({\bf k}+{\bf k}^{'},\omega+\omega^{'}) \times \nn
&& G^b_{ml}({\bf k}^{'},\omega^{'}) \Bigr] \hspace{0.2cm}\mbox{in the SU(2) theory}. 
\eqa
Here the spinon Green's function is $G^f_{\alpha \beta}({\bf k},\omega) = -i\int dt \sum_{\bf x} e^{i\omega t-i {\bf k} \cdot {\bf x}} < T[ f_{\alpha}({\bf x},t) f_{\beta}^\dagger(0,0) ] >$ and
the holon Green's function, $G^b({\bf k},\omega) = -i\int dt \sum_{\bf x} e^{i\omega t-i {\bf k} \cdot {\bf x}} < T[ b({\bf x},t) b^\dagger(0,0) ] >$.
They are the mean field Green's functions for the U(1) Hamiltonian.
The mean field Green's functions for the SU(2) Hamiltonian are $G^f_{\alpha\beta l m}({\bf k},\omega) = -i\int dt \sum_x e^{i\omega t-i {\bf k} \cdot {\bf x}} < T [ \psi_{\alpha l}({\bf x},t) \psi_{\beta m}^\dagger(0,0) ] >$
and $G^b_{lm}({\bf k},\omega) = -i\int dt \sum_x e^{i\omega t-i {\bf k} \cdot {\bf x}} < T [ b_{l}({\bf x},t) b_{m}^\dagger(0,0) ] >$ respectively.
The symbol $<$ $>$ refers to the finite temperature  ensemble average of an observable quantity $O$, $<O> \equiv \frac{1}{Z}{\rm tr}(e^{-\beta H}O)$.

The one electron removal spectral function, $I({\bf k}, \omega )$ is obtained from \cite{RANDERIA},
\bqa
I({\bf k},\omega) = -\frac{1}{\pi}{\rm Im G}({\bf k},\omega + i 0^+) f( \omega ), \\
\eqa
where $f(x)$ is the Fermi distribution function.
In the present study, we choose the Heisenberg coupling constant, $J=0.2$ $t$ and the hopping strength $t=0.44$ $eV$\cite{HYBERTSEN}.
Using the SU(2) theory , the predicted values of optimal hole doping rate $\delta$, pseudogap temperature $T^*$ and bose condensation temperature $T_c$ are $\delta = 0.13$, $T^* = 0.029 t$($148 K$) and $T_c = 0.021 t$ ($107.2 K$) respectively.
To compute the spectral function above we first evaluate the convolution integral of the holon and spinon Green's functions $G({\bf k}, \omega)$,  based on the  effective Hamiltonians of the spinon and holon sectors respectively.  
All of the computed results are based on the square lattice of $100 \times 100$ in momentum space which is found to be sufficient for numerical convergence.

Fig. 1 displays the momentum dependence of the computed spectral functions at optimal doping for the ranges of momentum from ${\bf k} = (0,0)$ to ${\bf k} = (\pi,\pi)$ and $(\pi, 0)$ to $( \pi, 0.4 \pi)$, by using the SU(2) theory. 
Variation of the predicted spectral peak positions with momentum is in qualitative agreement with the ARPES data\cite{SHEN}\cite{NORMAN}.
The predicted spectral functions at temperatures below $T_c$ are characterized by the presence of sharp peaks(quasiparticle peaks)  with shoulders(humps). 
Although not shown here, the U(1) slave-boson theory  also predicts similar structures with higher spectral peaks and lower humps.
In Fig.1 (a) the spectral peak position or the gap is seen to shift from a high value of $\sim 150 meV$ to a low value of $\sim 30 meV$ for the range of momentum from ${\bf k} = (0,0)$ to $(\pi, 0)$ at a temperature of $T= 0.004 t$($20.4 K$ with  $t = 0. 44 eV$\cite{HYBERTSEN}).  
Encouragingly the predicted value of the gap $\sim 30 meV$ at ${\bf k}  = (\pi, 0)$ is close to the value of spinon pairing gap $31 meV$ obtained from Eq.(\ref{eq:spinon_energy_gap}).
This indicates that the (leading energy) gap is now identified as the spinon pairing gap.
The formation of the spin singlet pairs(spinon pairs) is attributed to the opening of the pseudogap at the transition temperature(pseudogap temperature $T^*$).  
Although not completely shown in the figure, we find that the predicted gap size undergoes a continuous change(increase) as temperature is decreased from $T^*$ to a superconducting temperature $T_c$ and even to temperatures below $T_c$.  
This indicates that the observed leading edge gaps of the spectral functions in the superconducting state should have the same origin as the ones observed in the pseudogap phase. 
Thus the remaining problem is to explain how the sharply enhanced quasiparticle peaks occur in the superconducting state while such distinctively sharp peaks are not manifest above $T_c$, as is shown in Fig. 1(b).
However, the hump feature with no peak is observed in the normal state of $Bi2212$ system\cite{SHEN}\cite{NORMAN}, while there exists no such measured reports regarding other high $T_c$ cuprates such as $YBCO$.
Indeed, it will be of great importance to verify whether this hump feature is a universal nature of the normal states of the high $T_c$ cuprates.
We argue that the robustness of the sharp quasiparticle peaks is attributed to the enhancement of spinon pairing(spin singlet pair formation) owing to the influence of holon pair bosons present in the superconducting state.
In the following we will provide an explanation.
 
In order to see the role of the holon pair bosons on the appearance of the sharp peaks in the superconducting state, we first choose momenta only at the bottom of holon band in the convolution integral.
They are ${\bf k} = (0,0)$ for $b_1$ bosons, and ${\bf k} = (\pi, \pi)$ for $b_2$ bosons, which will allow for the formation of the holon pair bosons of the zero center of mass momentum ${\bf q}=(0,0)$ at temperatures below $T_c$\cite{BOTTOM}.
With such allowance of only the zero center of mass momentum for the holon pairs, sharp quasiparticle peaks with no shoulders are predicted to occur, as indicated by a dashed line in Fig. 2(a). 
On the other hand, with its removal such sharp peaks tend to be suppressed, as is shown by the solid line in the figure.
However, with the inclusion of all possible values of momenta including the zero center of mass momentum, i.e., ${\bf k} = (0, 0)$ and ${\bf k} = (\pi, \pi)$,  both the sharp peaks and the broad shoulders simultaneously appear, as is shown in Fig. 2 (b). 
Thus we argue that the sharp spectral peaks with leading edge gaps are caused by the enhancement of spinon pairing(spin singlet pair formation) by the presence of the holon pair bosons in the superconducting state. 
Although not displayed here, we find that physics is unchanged even with the U(1) mean field treatment. 
However some differences are observed in peak positions and heights. 
For completion the differences will be discussed shortly.
 
In Fig. 3 the doping dependence of spectral functions is displayed for underdoped, optimally doped and overdoped cases at a low temperature below $T_c$, $T = 0.004 t$ ($T = 20.4 K$ with $t = 0.44$ $eV$) with a choice of  ${\bf k} = (0. 8 \pi, 0)$(that is, close to ${\bf k} = (\pi, 0)$) for a qualitative comparison with observation.  
We find that the predicted spectral weight of the sharp quasiparticle peaks in the underdoped region decreases with decreasing hole concentration, showing agreement with the ARPES\cite{SHEN}\cite{HARRIS}. 
The peaks in the overdoped region, e.g., $\delta = 0.20$ are found to be higher than the ones in the underdoped region, in agreement with observation\cite{SHEN}\cite{HARRIS}, revealing that the spectral weight increases with hole concentration.  
 
Fig. 4 illustrates the role of the phase fluctuations of the spinon pairing order parameters on the spectral functions at an underdoping rate $\delta = 0.08$ by observing differences between the U(1) and SU(2) theories.  
It is reminded that the phase fluctuation effects of the order parameters are incorporated only in the SU(2) theory, but not in the U(1) mean field theory. 
The SU(2) theory predicted a lower peak with a higher shoulder(hump) compared to the U(1) theory. 
Thus phase fluctuations cause to enhance the shoulder by lowering the spectral weight of the quasiparticle peaks as a compensation. 
In addition we find that the predicted spectral position or the spin gap size is shifted to a larger binding energy compared to the U(1) result.
 
In the present study we examined how sharp quasiparticle peaks and broad shoulders in the one electron removal spectral functions are predicted based on both the SU(2) and U(1) slave-boson approaches to the t-J Hamiltonian. 
We found that the gap(or spectral peak position) undergoes a continuous change, manifesting a gradual increase with decreasing temperature from the pseudogap temperature $T^*$ to temperatures below the superconducting temperature $T_c$. 
This indicates that the origin of the observed leading edge gaps with the sharp(quasiparticle) peaks in the superconducting state is the same as the ones in the pseudogap phase.  
We showed that this gap is caused by the formation of spin singlet pairs(spinon pairs) that exist in both the pseudogap and superconducting phases. 
However the appearance of the distinctively sharp quasiparticle peaks below $T_c$ is attributed to the enhanced probability of spinon pairing(spin singlet pair formation) by the presence of the abundant holon pair bosons in the superconducting state, in contrast to the highly suppressed peaks predicted for the normal state. 
It is noted that broad shoulders with no peaks are observed in the normal state of $Bi2212$ system, while there exist no such reported measurements for other high $T_c$ cuprates such as $YBCO$. 
Thus it awaits further scrutiny to see whether such disappearance of the quasiparticle peaks is a common 'rule' for all the high $T_c$ cuprates.
It was shown that the spectral weight of the quasiparticle peaks increases with increasing hole concentration.
This trend is in complete agreement with the ARPES measurements. 
Finally we note from the comparison of the SU(2) and U(1) theories that the effects of phase fluctuations in the spinon paring order parameters result in the enhancement of shoulders by lowering the height of the sharp peaks('quasiparticle' peaks).
 
One(SHSS) of us acknowledges the generous supports of Korea Ministry of Education(BSRI-98 and 99) and the Center for Molecular Science at Korea Advanced Institute of Science and Technology.

\begin{minipage}[c]{9cm}
\begin{figure}
\vspace{0cm}
\epsfig{file=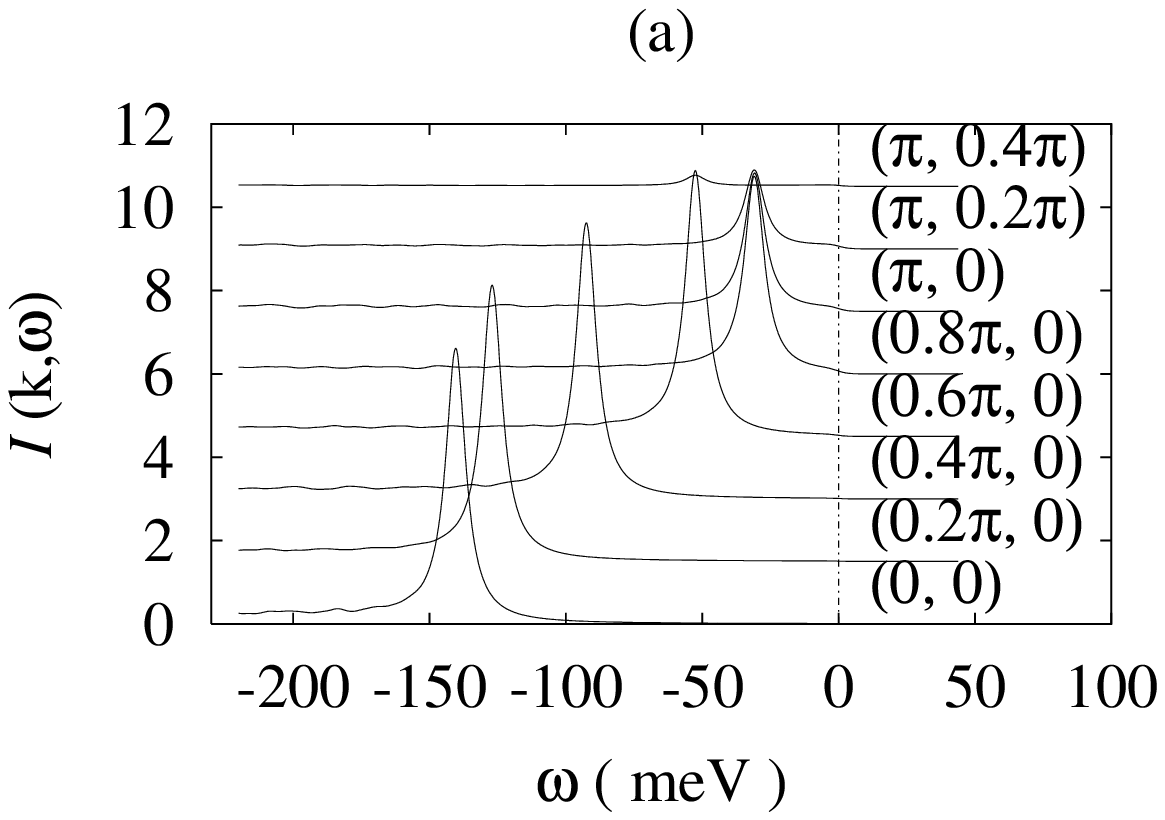, height=6.2cm, width=6.2cm}
\epsfig{file=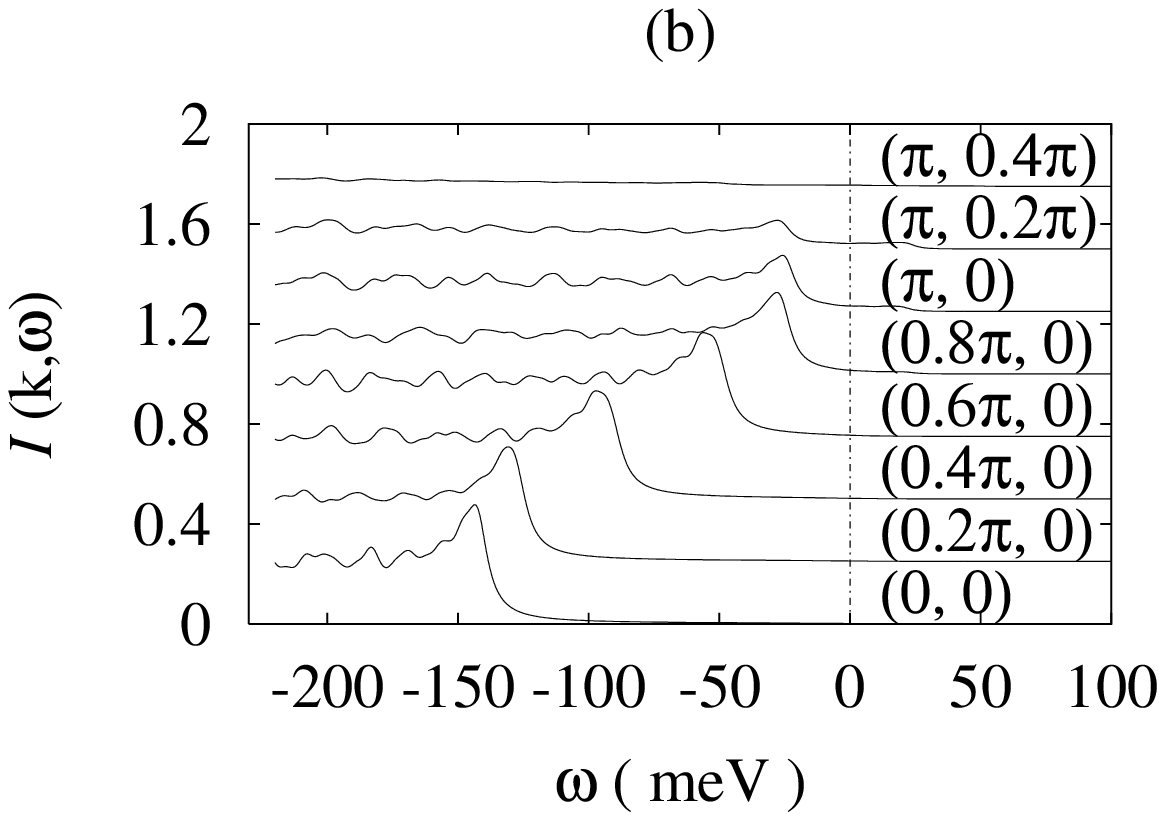, height=6.2cm, width=6.2cm}
\label{fig:1}
\caption{
Momentum dependence of spectral function $I({\bf k},\omega)$ by SU(2) theory($T_c=0.021t$($107.2 K$) at the optimal doping of $\delta=0.13$) for (a) $T=0.004t$($20.4 K$) and (b) $T=0.022t$($112.3 K$).
Two holon momenta $(0,0)$ and $(\pi, \pi)$ are excluded in the convolution integral for (b). 
}
\end{figure}
 \end{minipage}

\begin{minipage}[c]{9cm}
\begin{figure}
\vspace{0cm}
\epsfig{file=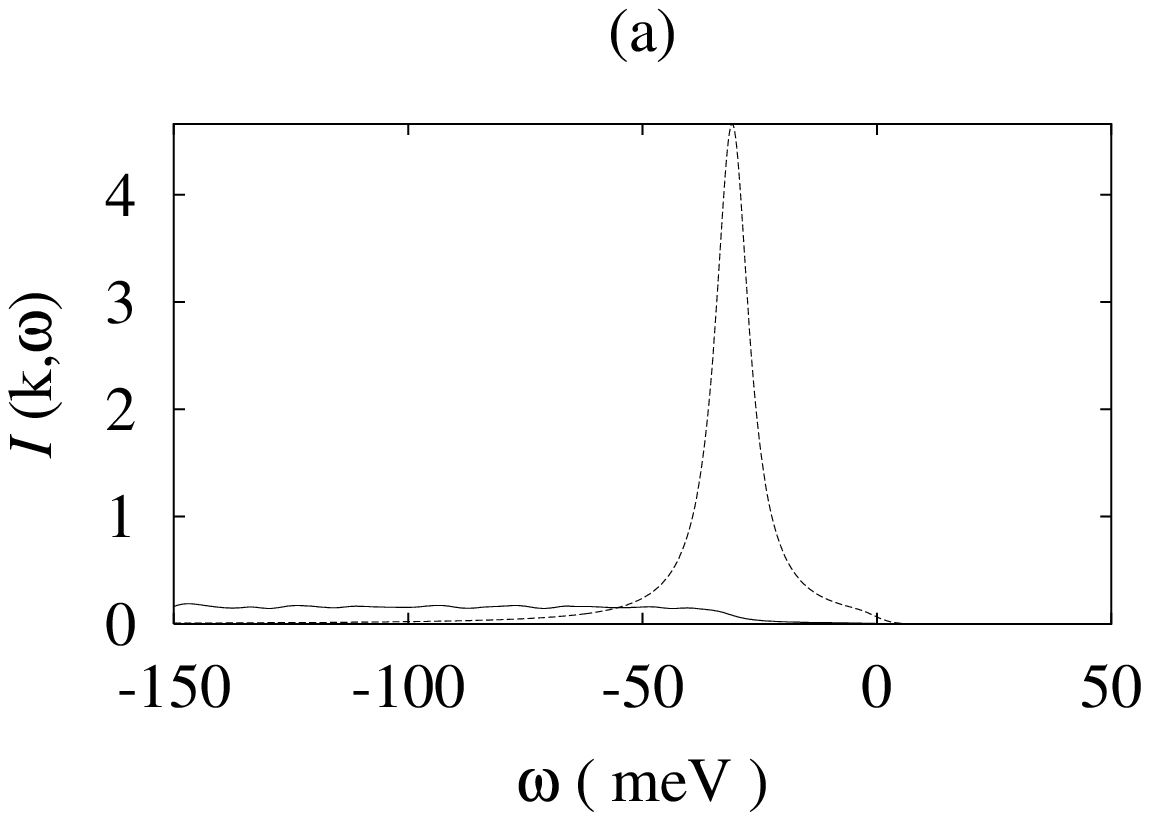, height=6.2cm, width=6.2cm}
\end{figure}
 \end{minipage}
\begin{minipage}[c]{9cm}
\begin{figure}
\epsfig{file=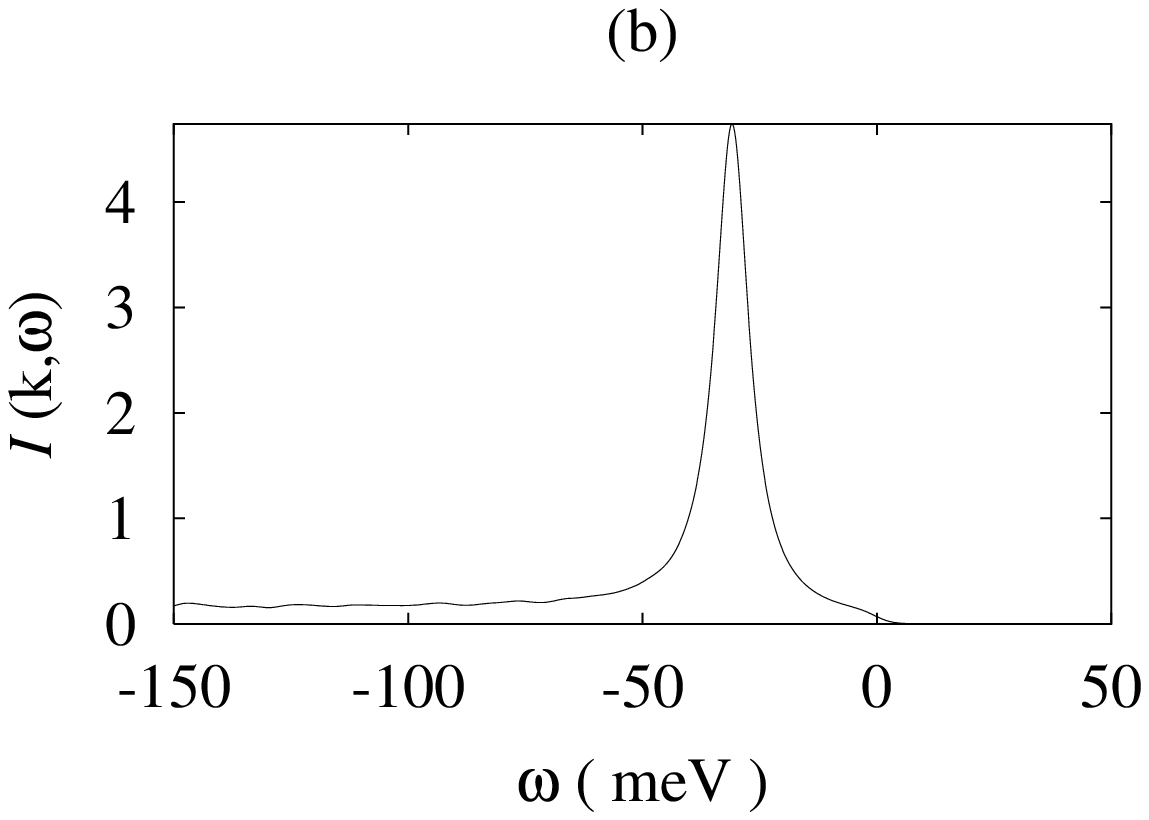, height=6.2cm, width=6.2cm}
\label{fig:2}
\caption{
Holon pair momentum dependence of spectral functions $I({\bf k},\omega)$ at $T= 0.004t$($20.4 K$) and ${\bf k}=(0.8 \pi , 0)$ with optimal doping($\delta=0.13$) : (a) The dashed line is $I({\bf k},\omega)$ contributed from the holon momentum ${\bf k}=(0,0)$ and $(\pi,\pi)$, and the solid line, $I({\bf k},\omega)$ with the exclusion of ${\bf k}=(0,0)$ and $(\pi,\pi)$. (b) $I({\bf k}, \omega)$ from the contributions of all possible holon momenta. 
}
\end{figure}
 \end{minipage}

\begin{minipage}[c]{9cm}
\begin{figure}
\vspace{0cm}
\epsfig{file=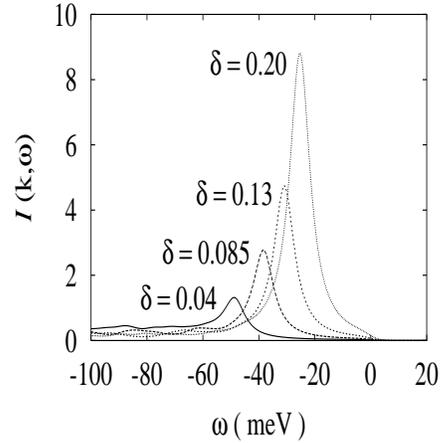, height=6.2cm, width=6.2cm}
\label{fig:3}
\caption{
Doping dependence of spectral functions $I({\bf k},\omega)$ at $T=0.004t$($20.4K$) and ${\bf k}=(0.8 \pi , 0)$ for underdoped($\delta=0.04$, $\delta=0.085$), optimal doping($\delta=0.13$) and overdoped($\delta=0.2$) rates.
}
\end{figure}
 \end{minipage}

\begin{minipage}[c]{9cm}
\begin{figure}
\vspace{0cm}
\epsfig{file=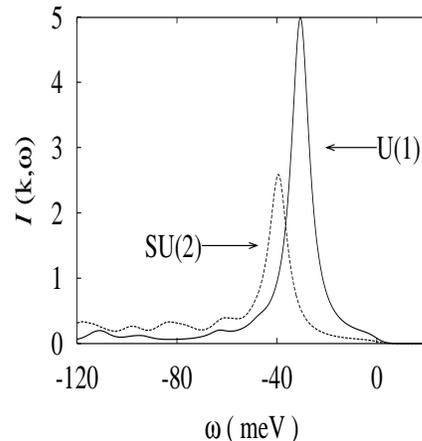, height=6.2cm, width=6.2cm}
\label{fig:4}
\caption{
U(1) and SU(2) spectral functions for $\delta=0.08$ at $T=0.004t$($20.4K$) and ${\bf k}=(0.8 \pi , 0)$.
}
\end{figure}
 \end{minipage}


\end{multicols}
\end{document}